# AMoRE: A search for neutrinoless double-beta decay of $^{100}$Mo using low-temperature molybdenum-containing crystal detectors

**Moo Hyun Lee**$^{a,b,*}$ on behalf of the AMoRE collaboration

$^a$ *Center for Underground Physics, Institute for Basic Science (IBS),*
*Daejeon, 34126, Korea*

$^b$ *IBS School, University of Science and Technology (UST),*
*Daejeon, 34113, Korea*
*E-mail*: mhlee@ibs.re.kr

ABSTRACT: The AMoRE is an experiment to search for neutrinoless double-beta decay of $^{100}$Mo in molybdate crystal scintillators using a cryogenic detection technique. The crystals are equipped with metallic magnetic calorimeter sensors that detect both phonon and photon signals at temperatures of a few tens of mK. Simultaneous measurements of thermal and scintillation signals produced by particle interactions in the crystals by MMC sensors provide high energy resolution and efficient particle discrimination. AMoRE-Pilot, an R&D phase with six $^{48depl}$Ca$^{100}$MoO$_4$ crystals and a total mass of ~1.9 kg in the final configuration, operated at the 700 m deep Yangyang underground laboratory (Y2L). After completion of the AMoRE-Pilot run at the end of 2018, AMoRE-I with a ~6 kg crystal array comprised of thirteen $^{48depl}$Ca$^{100}$MoO$_4$ and five Li$_2$$^{100}$MoO$_4$ crystals is currently being assembled and installed at Y2L. We have secured 110 kg of $^{100}$Mo-isotope-enriched MoO$_3$ powder for the production of crystals for the AMoRE-II phase, which will have ~200 kg of molybdate crystals and operate at Yemilab, a new underground laboratory located ~1,100 m deep in the Handeok iron mine that is currently being excavated and with a scheduled completion date of December 2020. AMoRE-II is expected to improve the upper limit on the effective Majorana neutrino mass to cover the entire inverted hierarchy neutrino mass region: 20-50 meV, in the case when no such decays are observed. Results from AMoRE-Pilot and progress of the preparations for AMoRE-I and AMoRE-II are presented.

---

$^*$ Corresponding author.

# Contents



## 1. Introduction

The Advanced Mo-based Rare process Experiment (AMoRE) is an experiment searching for neutrinoless double-beta (0νββ) decay using $^{100}$Mo isotopes in molybdate crystal scintillators operated as cryogenic detectors [1]. The 0νββ decay experiments aim to test the conservation of the lepton number, determine whether the neutrino is a Dirac or Majorana type particle, and estimate the absolute neutrino mass scale. If 0νββ decay is observed, it will be a sign of physics beyond the Standard Model [2]. If 0νββ decay is not observed, it will set limits on the half-life of the decay and the neutrino mass scale based on the total number of $^{100}$Mo nuclei in the detector's crystals. Of the naturally occurring candidate nuclides for 0νββ decay search experiments, $^{100}$Mo has some important advantages: a) high natural abundance (9.7%); b) promising theoretical predictions on the 0νββ decay probability [3,4]; c) relatively large $Q_{\beta\beta}$ value of 3034 keV, which is important because natural $\gamma$ backgrounds are small for energies above 2615 keV.

The molybdate crystals (enriched to more than 95% of $^{100}$Mo) in AMoRE are equipped with metallic magnetic calorimeter (MMC) sensors that detect both phonon and photon signals at temperatures of a few tens of mK. Simultaneous readout of thermal and scintillation signals produced by particle interactions in the crystals by the MMC sensors provides high energy resolution (a few keV) and efficient particle discrimination (α vs. β/γ) to reduce alpha-radiation-induced background signals [5]. The use of molybdate crystals as both the source and the detector of the decay provides close to 100% detection efficiency [6,7].

AMoRE has been planned to operate in three stages: The R&D stage, AMoRE-Pilot, precedes the next two phases, AMoRE-I and II, and is to learn about the operation of the cryostat and to understand the performance of the detector elements and determine the background levels from both internal and external sources. AMoRE-Pilot operated with six $^{48depl}$Ca$^{100}$MoO$_4$ crystal detectors (~1.9 kg mass) from August 2015 to December 2018 at the 700 m deep Yangyang underground laboratory (Y2L) in Korea. AMoRE-I, which has thirteen $^{48depl}$Ca$^{100}$MoO$_4$ and five Li$_2$$^{100}$MoO$_4$ crystal detectors (~6 kg mass) contained in the same cryostat used for the AMoRE-Pilot, will take data for about two years. AMoRE-II is the main phase of the project, with ~400 molybdate crystal scintillators (~100 kg of $^{100}$Mo in ~200 kg of crystals). Because the size of the AMoRE-II detector array and cryostat is too big to be installed in the existing Y2L space, which cannot be expanded, a new underground laboratory in Jeongseon, called Yemilab (after the name of the mountain where the laboratory is located) ~1100 m deep, is being constructed with completion expected by the end of 2020. Configurations of the AMoRE-Pilot, -I, and -II crystal arrays are shown in Fig. 1.

In the next three sections, the progress of AMoRE-Pilot, AMoRE-I, and AMoRE-II is presented. The last section will summarize the status of the AMoRE project.

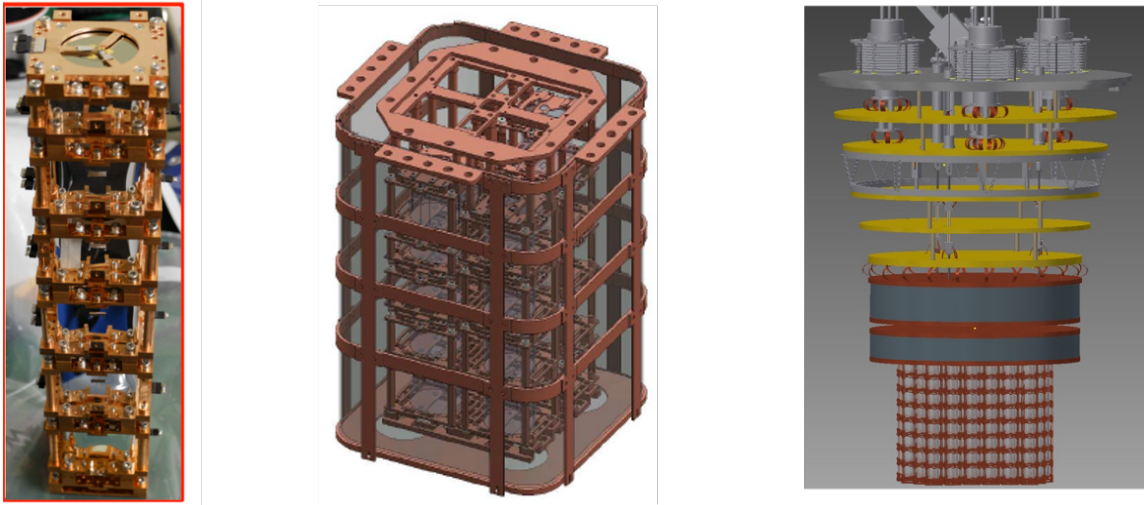

**Figure 1.** The left-most photo is the configuration of AMoRE-Pilot with six CaMoO$_4$ crystals (~1.9 kg in total). The middle figure is a schematic of the AMoRE-I configuration with thirteen CaMoO$_4$ crystals and five Li$_2$MoO$_4$ crystals (~6 kg in total). The right-most figure is a schematic of the AMoRE-II configuration with ~400 molybdate crystals (~200 kg in total) beneath lead shields and cold plates.

## 2. AMoRE-Pilot

AMoRE-Pilot has up to six calcium molybdate crystal scintillators ($^{48depl}Ca^{100}MoO_4$) depleted in $^{48}$Ca (less than 0.001% to reduce background from 2νββ decay of $^{48}$Ca [8]) and enriched in $^{100}$Mo (more than 95%) [9]. The crystals were produced by JSC "Fomos Materials" in collaboration with the AMoRE collaboration [10]. The operating temperature ranged between ~10 and ~30 mK. Each crystal was equipped with photon and phonon sensors to collect scintillation light and heat signals. Both sensor-types used metallic magnetic calorimeter (MMC) readouts [11]. The phonon collector is placed on the bottom surface of the crystal in the form of a gold patterned film with dimensions of 2 cm diameter with a maximum thickness of 400 nm that is evaporated onto the crystal surface. The scintillation light absorber is a 2-inch Ge wafer with three gold patterned films (5 mm diameter and 300 nm thickness) that are evaporated on the wafer [12]. Annealed gold-wires thermally connected the gold films to the MMC sensors. The Ge wafer and the crystal were isolated thermally as much as possible to avoid cross-talk between the phonon and photon channels. Each crystal was surrounded by a Vikuiti reflector to increase the light collection efficiency and mounted in a copper frame with Teflon support elements that are specially designed to reduce thermal effects from vibration. Two stages of vibration mitigation systems were employed to minimize the effects of vibrations on the signals from the operation of a pulse tube refrigerator used to operate a dry dilution refrigerator [13]. A 10 cm-thick ultra-low background lead shield placed over the detector assembly for radiation shield from the cryostat components around the cryostat attenuates background radiation from external sources [14].

There were seven AMoRE-Pilot operating runs with various improvements incorporated between runs, including: the installation of additional crystal elements; vibration reduction; plastic scintillator muon veto counter; replacement of high background detector components with lower ones; and more neutron shielding (boric-acid powder, borated rubber sheet, borated polyethylene (PE), and PE blocks). Each upgrade improved the overall detector performance in terms of energy resolution, particle discrimination, and background levels.

The AMoRE-Pilot detector array had six $^{48\text{depl}}$Ca$^{100}$MoO$_4$ scintillator elements with a mass of 1.887 kg (0.886 kg of $^{100}$Mo) as shown in Fig. 1 (left). Data from six photon and phonon channels were acquired by a 12 channel 18-bit FADC (NoticeKorea). Data were saved continuously in a DAQ server and triggered later off-line via the application of a Butterworth filter [15,16] or from a simple comparison of the integrated charges in two adjacent time windows. These trigger algorithms also enhanced the signal-to-noise ratio and improved energy resolution.

For the selection of beta decay events, various criteria were applied to improve the signal-to-background ratio in the region of interest (ROI) around the $Q_{\beta\beta}$ value of $^{100}$Mo. The selection criteria are based on operational status, event-by-event identification, and coincidence tagging. The muon veto system is used to reject signals induced by cosmic muons (the muon rejection efficiency was 87% – 92% depending on the crystal location).

The first result from AMoRE-Pilot from an energy spectrum for a 111 kg·day exposure is a half-life limit on the 0νββ decay of $^{100}$Mo of $T_{1/2}^{0\nu} > 9.5\times10^{22}$ yr [17]. For this measurement, the background level in the ROI was 0.55 counts·keV$^{-1}$·kg$^{-1}$·yr$^{-1}$. The lifetime limit is one order of magnitude below the NEMO-3 result ($1.1\times10^{24}$ yr) [18], but slightly better than the LUMINEU result ($7\times10^{22}$ yr) [19]. An analysis of the entire AMoRE-Pilot data exposure is in progress and should improve the experimental sensitivity [20].

## 3. AMoRE-I

The AMoRE-I experiment with ~6 kg of crystals, thirteen $^{48\text{depl}}$Ca$^{100}$MoO$_4$ and five Li$_2$$^{100}$MoO$_4$, was prepared and assembled in 2019. It is currently being cooled at Y2L using the same AMoRE-Pilot cryostat with some improvements. The thirteen $^{48\text{depl}}$Ca$^{100}$MoO$_4$ crystals include the six crystals used in the AMoRE-Pilot plus seven additional ones made by the same JSC "Fomos Materials" company [21,22]. The nine most recent crystals from "Fomos Materials", produced in 2015 to 2016, have a lower background level in the ROI (<10$^{-3}$ counts·keV$^{-1}$·kg$^{-1}$·yr$^{-1}$) than that for the previously produced ones (<10$^{-2}$ counts·keV$^{-1}$·kg$^{-1}$·yr$^{-1}$). The five Li$_2$$^{100}$MoO$_4$ crystals comprise four produced by Nikolaev Institute of Inorganic Chemistry (NIIC) and one produced at the Center for Underground Physics (CUP) [23-26]. The Li$_2$$^{100}$MoO$_4$ crystals are candidates for AMoRE-II, depending on their performance, background levels, and operating stability in AMoRE-I.

The detector assembly for AMoRE-I is shown in Fig. 1 (middle). The muon veto system has been improved by the addition of four more counters that cover gaps not covered in AMoRE-Pilot. The external lead shield is strengthened by adding a 5 cm thick lead layer around the cryostat. Additional neutron shields were installed including polyethylene (PE) blocks and layers of borated PE and boric acid. FADC modules that read out all the eighteen crystal detectors were added to the upgraded DAQ system together with a faster computer server with more disk space. The upgraded DAQ system is now equipped with optical links to the ADC modules and server that provide better isolation and a lower noise level. AMoRE-I is scheduled to begin in April 2020 and data taking should last for at least 2 years.

## 4. AMoRE-II

AMoRE-II will have ~200 kg of molybdate crystals ($^{48\text{depl}}$Ca$^{100}$MoO$_4$, Li$_2$$^{100}$MoO$_4$, Pb$^{100}$MoO$_4$, or Na$_2$$^{100}$Mo$_2$O$_7$) and will operate at Yemilab, a new, ~1,100 m deep underground laboratory in the Handeok iron mine that is currently being excavated with a scheduled completion date of December 2020. A schematic of the AMoRE-II detector configuration is shown in Fig. 1 (right).



We have secured 110 kg of $^{100}$Mo-enriched molybdenum in the form of MoO$_3$ powder from JSC/Isotope (Russia) for the production of crystals for AMoRE-II. Ultimately, AMoRE-II is expected to improve the upper limit on the effective Majorana neutrino mass to cover the entire inverted hierarchy neutrino mass region: 20-50 meV, in the case when no such decays are observed over the projected background level in the ROI of 10$^{-4}$ counts·keV$^{-1}$·kg$^{-1}$·yr$^{-1}$.

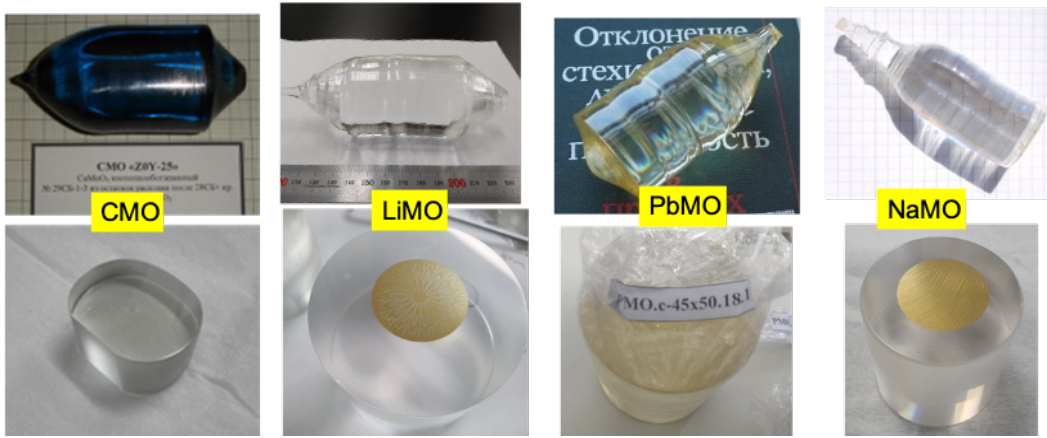

**Figure 2.** Photos of four molybdate candidate crystals (CaMoO$_4$, Li$_2$MoO$_4$, PbMoO$_4$, and Na$_2$Mo$_2$O$_7$) for the AMoRE-II experiment. The photos in the top row are ingots of each type and those in the bottom row are detector elements prepared for experiments and tests. The CaMoO$_4$ ingot color is dark blue because of an oxygen deficiency during the growth, and the shape of the detector element is not cylindrical because of difficulties in machining the side. The PbMoO$_4$ ingot and element show a yellowish color while those for Li$_2$MoO$_4$ and Na$_2$Mo$_2$O$_7$ are relatively transparent. More details are provided in the text.

Currently, $^{48\text{depl}}$Ca$^{100}$MoO$_4$, Li$_2$$^{100}$MoO$_4$, Pb$^{100}$MoO$_4$, and Na$_2$$^{100}$Mo$_2$O$_7$ are being considered as detector candidates for the AMoRE-II as shown in Fig. 2. The $^{48\text{depl}}$Ca$^{100}$MoO$_4$ (CMO) crystals used in AMoRE-Pilot and -I are excellent candidates with high light yields but have several disadvantages, especially the required depletion of $^{48}$Ca and a high melting temperature that make the production cost prohibitively high for a larger-scale experiment like AMoRE-II. Instead, we have been investigating many other molybdate crystals in terms of growth and characterizations related to cryogenic detection [27-29]. From these studies, the above-mentioned three types of crystals are candidates for the AMoRE-II. The Li$_2$$^{100}$MoO$_4$ (LiMO) crystal has been well studied and is easy to grow even though its hygroscopic property requires a well-controlled humidity environment during the preparation process. Backgrounds have already been characterized and found to be adequate for AMoRE-I. The Na$_2$$^{100}$Mo$_2$O$_7$ (NaMO) crystal developed by NIIC is being studied in collaboration with the AMoRE project [30,31]. Even though it has a light output that is comparable to that of CMO, it has some cleavage planes produced during crystal growth that complicates its cutting. Also, the NaMO crystal's background levels and performance at mK temperatures are currently under investigation at CUP. Pb$^{100}$MoO$_4$ (PbMO) crystals that incorporate ancient Pb to reduce background from $^{210}$Pb are interesting candidates [32,33]. Even though the light output produced by a charged particle radiation source is very high, we don't have enough experience using this with an MMC readout. Specifically, gold film evaporation has been a problem. It will be pursued in the future but with a lower priority than the testing of the other two candidate crystals (LiMO and NaMO). As shown in Fig. 2, we have samples of all the candidate crystals prepared for phonon-photon detection tests that will be completed soon.

For the growth of ~200 kg of low background molybdate crystals, we have investigated purification methods and crystal growing conditions at both NIIC and CUP [34]. Purification methods for CaCO$_3$ and MoO$_3$ have been presented recently [35-38]. For pure LiCO$_3$ production, a collaborative study between NIIC and CUP was started at the beginning of 2020 because the



pure LiCO$_3$ that was used for the production of existing LiMO crystals was from a stock produced in Russia a number of years ago and the remaining powder is only enough for 14 additional LiMO crystals. The remaining studies for NaMO and PbMO crystals are necessary to determine which crystal would be the best for a large-scale experiment like AMoRE-II. We plan on making a decision about which crystal will be used in AMoRE-II crystal by late mid-2020, after collecting data in AMoRE-I (LiMO) and from a dedicated small-scale cryogenic test setup at CUP (NaMO and PbMO) [39].

When the crystal type decision is made, we will have sufficient purified $^{100}$MoO$_3$ powder on hand. After making a contract with JSC/Isotope in late 2015, we secured 110 kg of $^{100}$Mo in the form of trioxide that is pure enough for the production of ~200 kg of molybdate crystals for the AMoRE-II as shown in Fig. 3 (left). Upon receiving the $^{100}$MoO$_3$ in ~15 kg batches, we assayed each batch with an inductively coupled plasma mass spectrometer (ICP-MS) and a high purity germanium (HPGe) detector array [40-43]. Figure 3 (right) shows the test setup for a 10 kg trioxide sample in an array of fourteen HPGe detectors used in a measurement that took about 3 months for the sample and another 3 months for background measurements.

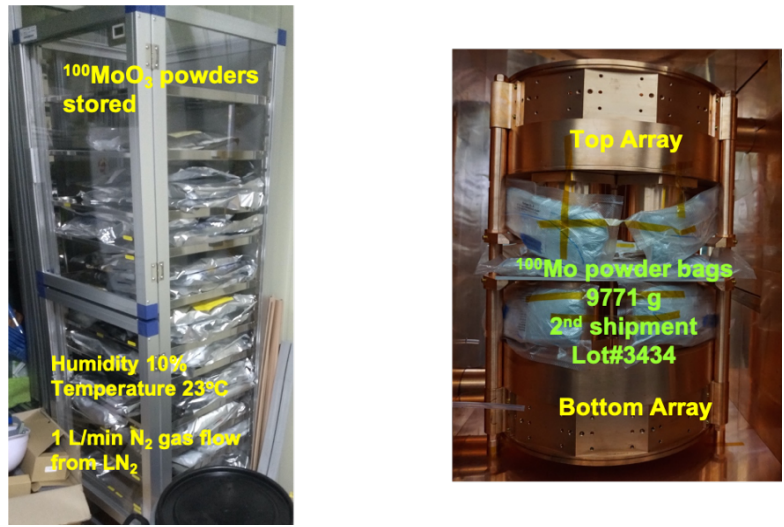

**Figure 3.** The left photo shows ~100 kg of $^{100}$Mo (in $^{100}$MoO$_3$ powder form) stored in a desiccator kept in relative humidity of ~10% and 23°C with a 1 L/min flow of boil-off gas from a liquid nitrogen dewar at Y2L. The right photo shows the measurement setup for a 10 kg sample of $^{100}$MoO$_3$ powder in an array of fourteen HPGe detectors (980% relative efficiency in total) to test contamination levels in powder slated to be used for growing ultra-pure molybdate crystals for AMoRE-II.

Figure 4 shows the excavation status of the Yemilab, which is scheduled to be completed by July 2020. As of late April 2020, the pit for the AMoRE-II and the remaining tunnels for other facilities and experiments are being excavated. The full construction of the Yemilab will be completed by the end of 2020 and will be ready for the construction of AMoRE-II.

## 5. Summary

AMoRE-Pilot, an R&D stage with six $^{48\text{depl}}$Ca$^{100}$MoO$_4$ crystal scintillator detectors (~1.9 kg) readout by MMCs and SQUIDs at mK temperatures, had seven successful runs between August 2017 and December 2019. After each run, various improvements were made for mechanical noise reduction, increasing the number of crystal elements, replacement of high background detector components with lower background ones, and installation of additional neutron shielding. The combined effect was improved energy resolution and detector performance. After a 111 kg·d



exposure with AMoRE-Pilot, an upper limit was set on the half-life of 0νββ decay of $^{100}$Mo as $T_{1/2}^{0\nu}$ > 9.5×10$^{22}$ yr at 90% C.L. which corresponds to an effective Majorana neutrino mass limit in the range of $m_{\beta\beta} \leq (1.2 - 2.1)$ eV. The background level in the ROI was measured to be 0.55 counts·keV$^{-1}$·kg$^{-1}$·yr$^{-1}$ in the data.

We plan to confirm a further reduction in the background level of the next phase experiment, AMoRE-I, which will use thirteen $^{48depl}$Ca$^{100}$MoO$_4$ and five Li$_2$$^{100}$MoO$_4$ crystal detectors for a total mass of ~6 kg. AMoRE-I is currently being cooled down to mK temperatures in the same cryostat that was used for AMoRE-Pilot in Y2L and will take data for at least two years, with an expanded muon veto array, improved DAQ system, and increased mass.

AMoRE-II will have a ~200 kg array of ultra-pure molybdate crystals and a reduced background level at the order of <10$^{-4}$ counts·keV$^{-1}$·kg$^{-1}$·yr$^{-1}$. To achieve this background level, the powder purification and crystal growth methods for various raw materials and crystals are being investigated both at NIIC and CUP. A 110 kg mass of $^{100}$Mo has already been secured for the production of molybdate crystals and is being kept in storage underground at Y2L except during the times used for purification processing at CUP.

AMoRE-II will operate at a new, ~1,100 m deep underground laboratory called Yemilab at the Handeok iron mine. The excavation of Yemilab started in March 2019 with completion expected by the end of 2020. AMoRE-II aims at improving the effective Majorana neutrino mass sensitivity down to the level of the inverted hierarchy of neutrino mass (20-50 meV).

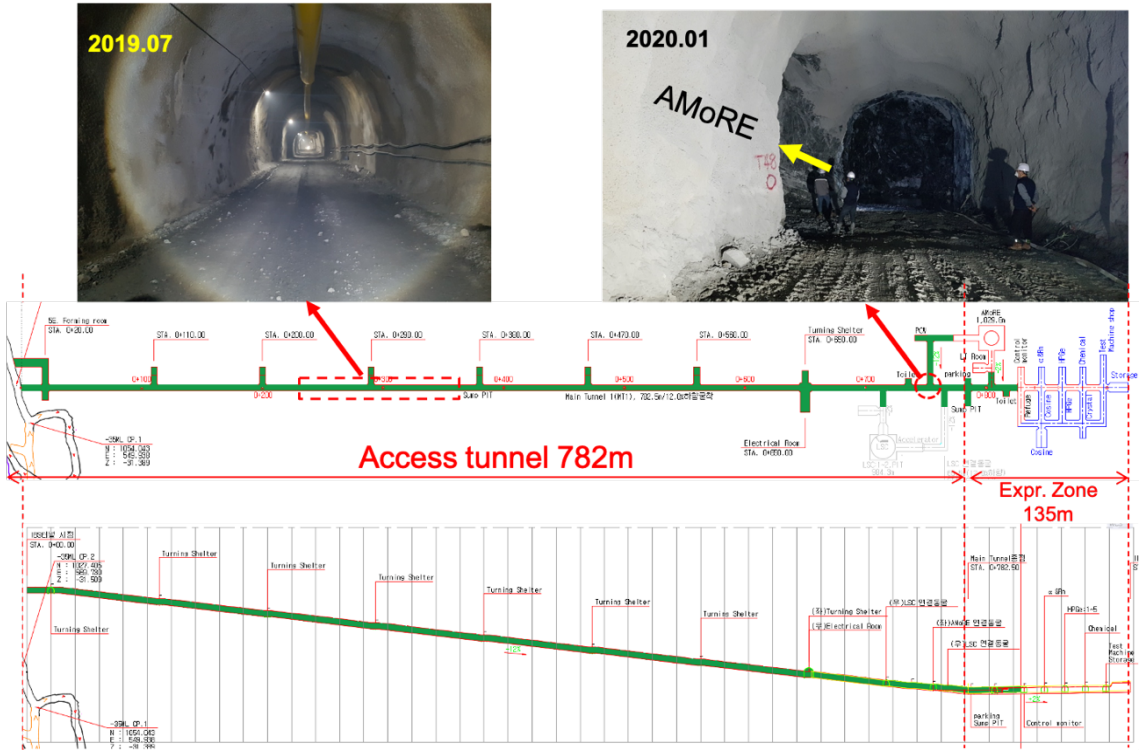

**Figure 4.** The Yemilab construction started in early 2019 at the Handeok mine in Jeongseon, Korea. The top left photo shows the view along ~200 m of the access tunnel taken in July 2019 and the top right photo shows the entrance to the AMoRE-II experimental hall on the left and main tunnel during excavation in January 2020. The bottom two figures show horizontal and vertical schematic views of the access tunnel and experimental areas with green color filled regions excavated in January 2020 and the empty red and blue color lined regions that will be excavated by summer 2020.

**Acknowledgments**

This research is supported by the following grants: IBSR016-D1 and IBS-R016-A2 from IBS, support in the framework of the Competitiveness Improvement Program NRNU MEPhI (contract No.02.a03.21.0005, 27.08.2013), the joint Ukraine-Republic of Korea R&D project "Properties of neutrino and weak interactions in double beta decay of $^{100}$Mo", the project "Investigations of rare nuclear processes" of the program of the National Academy of Sciences of Ukraine "Laboratory of young scientists" and the National Research Foundation of Korea (NRF) grant funded by the Korean government (MSIT) (No. NRF-2018K1A3A1A13087769).